\documentclass[prb,preprint,showpacs]{revtex4}

\usepackage{amsmath}

\begin{document}

\title{Acoustic lens design by genetic algorithms}
\author{A. H\aa kansson and J. S\'{a}nchez-Dehesa*}
\affiliation{Centro de Tecnolog\'{\i}a Nanofot\'{o}nica and Dpto. Ingenier\'{\i}a Electr%
\'{o}nica,\\
Universidad Polit\'{e}cnica de Valencia, E-46022 Valencia, Spain.}
\author{L. Sanchis}
\affiliation{Departamento de F\'{\i}sica Te\'{o}rica de la Materia Condensada, \\
Facultad de Ciencias,\\
Universidad Aut\'{o}noma de Madrid, E-28049 Madrid, Spain.}

\begin{abstract}
A survey of acoustic devices for focusing airborne sound is presented. We
introduce a new approach to design high quality acoustic lenses based on
arrays of cylindrical rigid scatterers in air. A population based stochastic
search algorithm is used in conjunction with the multiple scattering theory
to optimize a cluster of cylinders that focuses the sound in a prefixed
focal point. Various lenses of different sized clusters, for different
frequencies and with different focal lengths are presented. In general three
focusing phenomena are remarked, focusing due to refraction, diffraction and
focusing due to multiple scattering. The dependency on the frequency of the
incident sound and the focal distance is analyzed indicating that higher
frequencies and smaller focal distances favour larger amplifications in thin
lenses based on multiple scattering. Furthermore, the robustness of a
designed acoustic lens is studied, examining the focusing effect against
errors in the cylinders' positions and their radius.
\end{abstract}

\maketitle

\section{Introduction}

Sonic crystals (SCs) are materials consisting of periodic distributions of
acoustic scatterers in air. The periodicity in these materials give rise to
sonic band gaps, a range of frequencies for which sound propagation is
forbidden inside the crystal for some (pseudo gap) or all (full gap) wave
vectors. These stop bands have induced various application proposals such as
sound shields and acoustic filters \cite%
{Nature,SanchezPerezPRL,ZhenYePRL,CaballeroPRE99,RubioJOLT,CaballeroPRB01,WMRobertson,
LSanchisJASA,KushwahaAPL97,ZhenYePRE01,GoffauxAPL03,SanchezPerezAPL02}%
. Recently, the properties of SCs in the low frequency region, where the
dispersion relation is linear, have also been of interest for device
applications\cite{LSanchisPRB03,FCerveraPRL02,ZhenYePRE03,NGarciaPRE03}. For
these frequencies the SC behaves as a homogeneous medium with an acoustic
impedance\ larger than air. This property has allowed the construction of
two refractive devices, a convergent lens and an acoustic interferometer\cite%
{FCerveraPRL02}. On one hand, a comprehensive study of acoustic
interferometers has been reported by some of us\cite{LSanchisPRB03}. On the
other hand, various works\cite{ZhenYePRE03,NGarciaPRE03,
AAKrokhinPRL03,CHKuo04,AHakansson04} have been published about focusing
sound and image formation in the light of the seminal work by Cervera 
\textit{et al.}\cite{FCerveraPRL02}. Different SC lenses have been proposed
and a discussion has aroused about whether refraction or diffraction is the
dominant mechanism for focalization\cite{NGarciaPRE03,AHakansson04}.
Focusing sound can also be obtained by using time-reversal techniques\cite%
{MFink00,MFink99}, but the discussion of these techniques is out of the
scope of the present work.

The goal of this paper is to answer the following question: Which is the
optimal SC structure that acts as an acoustic lens focusing the sound at a
predetermined focal distance? From this point of view we will discuss what
is the physics that control the focusing process in several acoustic lenses,
which were optimized for different frequencies and focal distances. Our
approach to the problem consists in using a population based stochastic
search algorithm, specifically a genetic algorithm (GA) \cite{HollandBook}
in conjunction with the multiple scattering theory (MST) \cite%
{VTwerskyJASA51}. This approach has been previously used by us to design a
spot-size converter in the field of two-dimensional photonic crystals\cite%
{LSanchis04}. GAs have been proved to be very efficient, especially in
optimization problems having a large set of discrete parameters. To speed-up
the stochastic search, here we have implemented a symmetric multiple
scattering theory (SMST) that decreases the calculation time needed for
predicting the scattered acoustic wave from a symmetric system.

\bigskip

\section{Method of calculation}

\subsection{Genetic Algorithm}

Evolutionary computation is a family of algorithms that are related with the
optimization process used by nature itself, the evolution. The evolution is
very powerful in adapting individuals to a given environment, and is able to
tackle enormous complex problems with fairly simple means. Throughout
generations the individuals' chromosomes (the genotypes) are mixed and new
individuals with different characteristics are born. Those individuals that
adapt better to the environment have the best chance of survival and hence
give birth to more offspring creating a new generation more fit for survival
than the previous one.

We here use one of the most popular algorithms from this family, a
simple-GA, introduced by Holland \cite{HollandBook}. The GAs are population
based stochastic search algorithms normally used to solve discrete problems.
Their functionality is based on their ability to find small parts of the
individual (the solution) that reflects a good quality in the result, these
fractions are called building blocks (BBs). In other words, the GA creates,
mixes and finally uses these BBs to build a global optimal solution.
Unfortunately this goal is not always met. If the global optimum consists of
large sized BBs, the GA will have problems of constructing them because
high-order BBs are very difficult to grow. These problems are referred to as
GA-hard problems\cite{GoldbergBook89}. Nevertheless, this is not of great
concern here since this work concentrates on the method itself and not on
the algorithm. At this point, we are pleased by finding good solutions
within the boundaries of the optimal.

The simple-GA works with three operators; selection, crossover and mutation
that are iteratively applied to a population of individuals. The selection
operator culls the population selecting better solutions over worse for
mating. The mating act, which is directed by the crossover operator, mixes
and constructs the BBs from two or more individuals creating new offspring.
The mutation operator makes it possible for the offspring to possess new BBs
that none of the parents have and is applied before passing into the next
generation (i.e., the next iteration). The individual is represented by a
chromosome (a digital string) that is put together by a number of binary
genes coding the genotype. Each gene corresponds to one specific part of the
phenotype and the corresponding value of the gene is called allele. The
genotype represents one and only one phenotype that here is a cluster of
rigid cylinders.

Our goal is to find which cluster focuses the sound most efficiently in a
fixed predetermined focal point, under a chosen constrain or boundary
condition. This cluster of cylinders can be seen as a transparent acoustic
lens device. To quantitatively judge the quality of such a device, a real
number is assigned to each cluster. This number is referred to as the
fitness of the solution and is here calculated as the pressure in the focus
if not mentioned others. One lens is better than another if its fitness is
superior

To predict the fitness we need to calculate the scattered pressure field
from a cluster of cylindrical scatterers. We here apply the Multiple
Scattering Theory (MST) that earlier has been used by some of us with good
agreement between theory and experiments \cite{LSanchisPRB03}. Because the
search space of the problem is very large, the GA needs to calculate the
fitness for numerous individuals before converging. For example, a genotype
with 50 one-bit genes could represent up to $\sim 10^{15}$ different
phenotypes. Thankfully the GA does not get near this number of fitness
calculations, but it is crucial to do the fitness calculation as fast as
possible. In order to speed\ up the fitness calculation, we apply a simple
symmetric condition on the space of solutions. The scattered pressure field
from a symmetric system has a mirrored symmetry with respect to the central
axis. In other words, the field scattered from a cylinder off the symmetry
axis is identical, only mirror reversed, as the field scattered from its
symmetric associate. This means if using MST a lot of computational
resources will be used to calculate the same scattered field twice for
almost every cylinders in the symmetric cluster. By including this mirror
symmetry in the MST the fitness calculation and the GA search are
considerably accelerated.

\subsection{Symmetric Multiple Scattering Theory}

The complete self-consistent procedure including all order of multiple
scattering\cite{Ishimaru} that follows the seminal work of Twersky\cite%
{VTwerskyJASA51} has been reported by Chen and Ye \cite{ZhenYePRE01} to
study the transmission spectra of large structures. The SMST trimmed to
handle symmetric systems will be introduced below.

Consider a cluster of N scatterers located at the positions $\overset{%
\longrightarrow }{R}_{\beta }\left( \beta =1,2,...,N\right) $, all placed
above or on the symmetry axis (the x-axis in Fig. 1). The complete symmetric
crystal is represented if each cylinder placed off the \textit{x}-axis is
copied and placed at the same \textit{x}-position and with the same, but
negative, $y$ coordinate. All cylinders placed below the symmetry axis are
indicated with a \textquotedblleft +\textquotedblright\ in the index (see $%
\alpha ^{+}$ in Fig. 1).

If a symmetric external wave $P_{sym}^{ext}$ with temporal dependence $%
e^{-i\omega t}$ impinges the cluster, the total field around the cylinder $%
\alpha $ is a superposition of the external field and the radiation
scattered by the cluster:

\begin{equation}
\bigskip P_{\alpha }(x,y)=P_{sym}^{ext}(x,y)+\underset{\alpha \neq \beta }{%
\overset{N}{\sum }}\left( P_{\beta }^{scatt}(x,y)+(1-\delta _{y})P_{\beta
^{+}}^{scatt}(x,y)\right) ,\qquad  \label{eq1}
\end{equation}%
where $\delta _{y}=0$ if $y_{\beta }\neq 0$, and $\delta _{y}=1$ if $%
y_{\beta }=0$. $P_{\beta }^{scatt}$ and $P_{\beta ^{+}}^{scatt}$ is the
field scattered by cylinder $\beta $ and its mirror image $\beta ^{+}$
respectively.

These four fields can be expanded into series Bessel functions. The total
incident wave to the $\alpha $ cylinder, which consists of a superposition
of the external incident wave and the scattered waves from all the other
cylinders in the cluster, is expressed in terms of the Bessel function of
the first kind as

\begin{equation}
P_{\alpha }(x,y)=\underset{l=-\infty }{\overset{\infty }{\sum }}(B_{\alpha
})_{l}J_{l}(r_{\alpha }\kappa )e^{il\theta _{\alpha }}  \label{eq1.1}
\end{equation}%
\bigskip where $\kappa $ is the wavenumber calculated as $\kappa =\frac{c}{%
\omega }$, where $c$ is the sound velocity in air, i.e. 340 m/s.

The external incident plane wave, impinging to cylinder $\alpha $ follows
the same expansion principle but with the multipole coefficients $(S_{\alpha
})_{l}$. The scattered waves from cylinder $\beta $ and $\beta ^{+}$ are
expressed in a series of outgoing Hankel functions rather than regular wave
functions:

\begin{equation}
P_{\beta }^{scatt}(x,y)=\underset{l=-\infty }{\overset{\infty }{\sum }}%
(A_{\beta })_{l}H_{l}(r_{\beta }\kappa )e^{il\theta _{\beta }}  \label{eq1.2}
\end{equation}

Using the multipole coefficients $(B_{\alpha })_{l}$ , $(S_{\alpha })_{l}$, $%
(A_{\beta })_{l}$ and $(A_{\beta ^{+}})_{l}$ for $P_{\alpha }$, $P_{\alpha
}^{ext}$, $P_{\beta }^{scatt}$ and $P_{\beta ^{+}}^{scatt}$ respectively,
the expression above can be cast into the following relation between
coefficients:

\begin{equation}
\left( B_{\alpha }\right) _{l}=\left( S_{\alpha }\right) _{l}+\underset{%
\beta =1}{\overset{N}{\sum }}\underset{l^{\prime }=-\infty }{\overset{%
l^{\prime }=\infty }{\sum }}\left( \left( G_{\beta \alpha }\right)
_{ll^{\prime }}\left( A_{\beta }\right) _{l^{\prime }}+\left( 1-\delta
_{y}\right) \left( G_{\beta ^{+}\alpha }\right) _{ll^{\prime }}\left(
A_{\beta ^{+}}\right) _{l^{\prime }}\right)  \label{eq2}
\end{equation}

$G_{\beta \alpha }$ and $G_{\beta ^{+}\alpha }$ being the propagator from
cylinder $\beta $ and $\beta ^{+}$ to $\alpha $ respectively, whose
components are

\begin{eqnarray}
\left( G_{\beta \alpha }\right) _{ll^{\prime }} &=&\left( 1-\delta _{\alpha
\beta }\right) e^{i(l^{\prime }-l)\theta _{\beta \alpha }}H_{l^{\prime
}-l}^{(1)}(\kappa r_{\beta \alpha })  \label{eq3} \\
\left( G_{\beta ^{+}\alpha }\right) _{ll^{\prime }} &=&\left( 1-\delta
_{y}\right) e^{i(l^{\prime }-l)\theta _{\beta ^{+}\alpha }}H_{l^{\prime
}-l}^{(1)}(\kappa r_{\beta ^{+}\alpha })  \notag
\end{eqnarray}

where $\delta _{\alpha \beta }$ is the Kronecker delta.

Notice that the coefficients $S_{\alpha }$ are known, but $B_{\alpha }$ , $%
A_{\beta }$ and $A_{\beta ^{+}}$ are not. The boundary conditions at a rigid
cylinders surface states that the normal pressure gradient equals zero. This
equation relates $B_{\alpha }$ with $A_{\beta }$ and the symmetric condition
relates $A_{\beta }$ with $A_{\beta ^{+}}$. The rigid cylinder approximation
has been demonstrated to reproduce fairly well experiments made of hollow
aluminum cylinders\cite{LSanchisPRB03}. Now the transition matrix can be
calculated and gives the proper transformation of the total incident field
to cylinder $\alpha $ expressed by the $B_{\beta }$ coefficients, to the
scattered field expressed using the $A_{\alpha }$ coefficients:

\begin{equation}
\left( t_{\alpha }\right) _{ll^{\prime }}=\frac{J_{l-1}\left( \kappa
r_{\alpha }\right) -J_{l+1}\left( \kappa r_{\alpha }\right) }{%
H_{l+1}^{(1)}\left( \kappa r_{\alpha }\right) -H_{l-1}^{(1)}\left( \kappa
r_{\alpha }\right) }\delta _{ll^{\prime }}  \label{eq4}
\end{equation}%
,and the mirror symmetry $(\theta _{\beta ^{+}}=-\theta _{\beta },r_{\beta
^{+}}=r_{\beta })$ relates $A_{\beta }$ and $A_{\beta ^{+}}$

\begin{gather}
P_{\alpha ^{+}}=\overset{l^{\prime }=\infty }{\underset{l^{\prime }=-\infty }%
{\sum }}\left( A_{\beta ^{+}}\right) _{l^{\prime }}H_{l^{\prime }}\left(
\kappa r_{\beta ^{+}}\right) e^{il^{\prime }\theta _{\beta ^{+}}}=\overset{%
l^{\prime }=\infty }{\underset{l^{\prime }=-\infty }{\sum }}\left( A_{\beta
}\right) _{l^{\prime }}H_{l^{\prime }}\left( \kappa r_{\beta ^{+}}\right)
e^{il^{\prime }(-\theta _{\beta ^{+}})}=  \notag \\
\overset{l^{\prime }=\infty }{=\underset{l^{\prime }=-\infty }{\sum }}\left(
A_{\beta }\right) _{-l^{\prime }}H_{-l^{\prime }}\left( \kappa r_{\beta
^{+}}\right) e^{il^{\prime }\theta _{\beta ^{+}}}=\overset{l^{\prime
}=\infty }{\underset{l^{\prime }=-\infty }{\sum }}\left( A_{\beta }\right)
_{-l^{\prime }}(-)^{l^{\prime }}H_{l^{\prime }}\left( \kappa r_{\beta
^{+}}\right) e^{il^{\prime }\theta _{\beta ^{+}}}  \label{eq5} \\
\left( A_{\beta ^{+}}\right) _{l^{\prime }}=\left( A_{\beta }\right)
_{-l^{\prime }}(-)^{l^{\prime }}  \label{eq6}
\end{gather}

Introducing the coefficients from Eq. \ref{eq4} and \ref{eq6} in Eq. \ref%
{eq2} will result in

\begin{equation}
\underset{\beta =1}{\overset{N}{\sum }}\underset{l^{\prime }=-\infty }{%
\overset{l^{\prime }=\infty }{\sum }}\left( A_{\beta }\right) _{l^{\prime
}}\left( \delta _{\alpha \beta }\delta _{ll^{\prime }}-\left( t_{\alpha
}G_{\beta \alpha }\right) _{ll^{\prime }}+(-)^{l^{\prime }}\left( t_{\alpha
}G_{\beta ^{+}\alpha }\right) _{l-l^{\prime }}\right) =\left( t_{\alpha
}S_{\alpha }\right) _{l}  \label{eq7}
\end{equation}

\bigskip By truncating the angular momentum within $\left\vert l^{\prime
}\right\vert \leq l_{\max }$, Eq. \ref{eq7} reduces to a linear equation
where the dimension of the relevant matrix is $N(2l_{\max }+1)\times
N(2l_{\max }+1)$. Note that $N$ is not the total number of cylinders, only
the cylinders placed on or above the \textit{x}-axis. For a system of 100
cylinders truncated to $l_{\max }=3$, the CPU-time needed to calculate the $%
A $ coefficients is reduced by a factor of 4 by using the SMST instead of
the standard MST. All results presented are calculated with $l_{\max }=3.$%
This approach is supported by the fact that no error higher than 1\% was
observed in any of the following calculations of the pressure field when
including higher angular momentums in the series expansion. The coefficients
in the transfer matrix give a good estimation of the convergence of the
method. Here, the most significant matrix coefficient with an index higher
than three is of the order \ $\sim 1\times 10^{-5}$. In all optimization
processes the angular momentum was truncated at $l_{\max }=1$ due to
computational resources. Since this will not guarantee that the simulated
fitness is within an proper error, a local search was later done with $%
l_{\max }=3$. The local search was done using a GA with a much smaller
population where one individual in the initial population correspond to the
one with maximal fitness in the search with $l_{\max }=1.$ To confirm the
quality of this two-step method a GA-optimization was done using $l_{\max
}=3 $ directly, with the result almost identical to the result received with
the two-step optimization method, but significantly more time consuming.

Figure 2 illustrates a typical GA-run using the SMST for designing
an acoustic lens. The problem is coded on 100 one-bit genes. As can be seen
in the figure the algorithm has converged after approximately 50,000 fitness
calculations, this equals one $10^{-30}$th part of the total number of
phenotypes.

\bigskip

\section{Results and discussion}

\subsection{Thick lenses}

Guided by recent works \cite%
{FCerveraPRL02,ZhenYePRE03,NGarciaPRE03,CHKuo04,AHakansson04}, where a
discussion has aroused concerning the shape of SCs devices for focusing
sound, we will here as the first step in our study treat the problem of
finding the optimal shape of a SC cluster for different focal distances and
for different symmetry constrains.

Firstly the GA was implemented to find the optimal SC cluster, symmetric
with respect to both axes, that focuses the sound on the $x$-axis for a
fixed focal distance. To do this, the device's genotype was implemented
using 10 two-bit genes, where each gene corresponds to a different layer in
the cluster. The first gene on the chromosome corresponds to the central
layer at $x$ = 0 and the four possible alleles of the gene (00, 01, 10, 11)
code the width of this layer, 20, 18, 16 or 14 cylinders respectively. This
can be interpreted as (00) keeps the width (20 cylinders for $x=0$), (01),
(10) and (11) remove one, two and three cylinder respectively, at each
extreme (leaving the 18, 16 or 14 cylinders for $x=0$). The next gene on the
chromosome follows this rule and is related to the two layers next to the
central one. Since the crystal is symmetric the same gene code both
neighboring layers. Each following gene on the chromosome relates to the
next neighboring layer and the allele tells us if this layer will decrease
in width (01, 10, 11) or be kept constant (00) with respect to the previous
layer.

We have considered a hexagonal array of rigid cylinder with radius r = 2 cm
and a lattice constant $a$ = 6.7 cm. Systems having these parameters have
been studied earlier both experimentally and theoretically in Refs. \cite%
{LSanchisPRB03,FCerveraPRL02,ZhenYePRE03,NGarciaPRE03}. We also consider an
incident sound plane wave traveling in the positive $x$-direction at 1700 Hz
. This wave is perfectly symmetric with respect to the \textit{x}-axis,
which is the only criteria for using the SMST and it will be assumed as
incident wave in all the following calculations if not mentioned others.
Four different lenses for four different focal distances were designed using
the GA. The same boundary conditions were applied, i.e. the central layers
maximal width is 20 cylinders and the maximum number of layers equals 19 (18
layer off-axis plus the central layer on-axis). The results are shown in
Fig. 3, where the amplification [in decibels (dB)] was calculated as

\bigskip

\begin{equation}
P(dB)=20\log \frac{\left| P(x_{f},y_{f})\right| }{\left|
P_{0}(x_{f},y_{f})\right| },
\end{equation}
where $(x_{f},y_{f})$ defines the focal point, and $|P_{0}|$ equals the
pressure measured in free space that here equals 1 for the plane wave.

The amplification in the optimized focal point for the four structures vary
from 8.1 dB to 8.6 dB. The highest gain was achieved for the smallest focal
length. A noteworthy observation is that the structure increases in width at
the same time as the curvature of the surface decrease with longer focal
length, a similar characteristic that for optical lenses and should be the
case for homogenous refractive devices. This can be considered as a support
of the work presented in Refs. \cite{FCerveraPRL02} and \cite{ZhenYePRE03},
i.e. the SC is seen as a homogeneous material and the dominant effect in
focalization is due to refraction. If comparing the designed structure in
Fig. 3c with the SC used in the experimental set-up in Ref. \cite%
{FCerveraPRL02} one can clearly observe a close similarity in the curvature
of the different surfaces.

The boundary conditions constraining the pool of solutions are very strong
in this first approach simplifying the optimization process. Drawn by the
determination to increase the amplification of the device we lifted some of
these constrains. Figure 4 shows three different devices design by the GA to
maximize the focalization at \textit{x} = 1.05 m (the same focal distance as
in Fig. 3c). The upper crystal has the most restrictions and the lower the
least. This means that the number of possible solutions increases with the
freedom of the GA and results in a harder problem to solve. The device in
Fig. 4a is the result from a genotype including 10 three-bit genes, where
each gene is related to the phenotype in a similar way as earlier. Here the
three--bit gene corresponds to eight different alleles that directly code
the number of cylinders in each layer; For the nine central layers in the
crystal: from 20 or 19 cylinders for the allele (000) down to 6 or 5 for
(111); For the ten outer layers (five to the left and five to the right) in
the crystal: from 14 or 13 cylinders for the allele (000) down to no
cylinders for (111). The boundary condition is set in such a way that the
crystal from Fig. 3c is included in the search space and hence mark a minima
for the expected optimization result. The increase in the genotype size
gives rise to an increase in the total number of possible solutions from $%
2^{20}$ $(\sim 10^{6})$ in Fig. 3c to $2^{30}$ $(\sim 10^{9})$ in 4a. The
resulting device is of the same size as in Fig. 3c but now the crystal is
not lens-shaped but formed out of three thinner sections. As expected the
amplification increases, from 8.1 dB to 8.3 dB.

In the next modification of the boundary condition we leave the homogeneous
SC cluster and give freedom to the algorithm to introduce point-defects in
the cluster. We here leave the homogenization discussion and let the GA play
with the complex procedure of multiple scattering inside the cluster of
scatterers, though the double symmetric condition still holds. The genotype
now equals 100 one-bit genes where each one-bit gene represents the presence
(1) or the absence (0) of one ($x=0$ and $y=0$), two ($x=0$ or $y=0$) or
four ($x\neq 0$ and $y\neq 0$) cylinder, due to the symmetry condition. The
search space is increased from $\sim 10^{9}$ to $2^{100}$ ($\sim 10^{30})$
possible solutions and the optimized amplification is increased from 8.3 dB
to 8.5 dB. Finally, in the last run, the symmetry with respect to the y-axis
is removed and the number of one-bit genes in the chromosome is increased to
200 which result in the gigantic number of possible solution equal $2^{200}$
($\sim 10^{60}$). Fig. 4c shows that an enhancement of sound focalization is
obtained.

In principle, one would expect that giving more degrees of freedom in the
inverse design process and so increasing the search space will increase the
quality in the result after the optimization, but this can be a misleading
assumption. Even though the results presented here follow this supposition,
it is interesting to compare the results in Fig. 4a and Fig. 4b.The
qualities differ only 0.2 dB while the total number of possible solutions
increases from $\sim 10^{9}$ to $\sim 10^{30}$. Since the number
of solutions increase exponentially (base 2) with the size of the chromosome
it is very important to define the genotype in a good way keeping down the
size of the problem and coding the BB in a way that minimizes their size.
Here we have kept the GA-parameters constant during the different runs. The
result in general can be enhanced trimming the GA-parameters to handle
larger and harder problems. When problems of this size need to be tackled, a
more advanced algorithm should be at hand, e.g. cGA (compaq GA), fmGA (fast
messy GA), BOA (Bayesian Optimization Algorithm) \cite{GoldbergBook02} that
are capable of dynamically finding larger sized BBs.

\subsection{Thin Lenses with Flat Surfaces}

\subsubsection{Five Layers lenses}

Following the path of the previous analysis we now continue our study using
a reduced SC framework consisting of five layers and 98 cylinders, compared
to earlier nineteen layers and 371 cylinders. The degree of freedom is set
at the previous highest level using 50 two-bit genes, each gene coding one
cylinder. The four possible alleles of the gene are representing three
different radius of the cylinders; (01): 1cm, (10): r=1.5 cm and (11): r=2
cm and (00): no cylinder present.

Figures 5 and 6 illustrate eight different devices for focusing the sound at
eight different distances along the symmetry axis. These flat lenses
represented by considerably few cylinders is perfectly competitive with the
thick lenses earlier presented (e.g. compare Fig. 4c. with Fig. 6c, two
lenses with the same focal distance). Although the really impacting results
are for short focal lengths were the amplification reaches up to 10.3 dB.
For longer focal lengths presented in Fig. 6a homogenous cluster of
cylinders, where no point defect is present, can be observed as a central
piece of the device. This cluster is more or less formed as a trapezoid
where the shorter of the two parallel sides always is closer to the focus.
That the cluster is homogeneous indicate that the focusing mechanism for
this part is mostly due to refraction or diffraction of the cluster. We can
clearly see in Fig. 6e that for this large focal distance (2m) the flat
lens, almost free of defects, works because of the diffraction at the
borders of the cluster and the focal-spot is almost completely lost.

Figure 7 displays the eight results put together in order to show how the
amplification decreases with an increase in the focal distance. This is a
surprising result since lenses working due to refraction and follow
Lensmaker's formula\cite{CHKuo04}, need to be very thick to achieve high
F-numbers. The F-number, calculated as the focal length divided by the
diameter of the lens, is for the structure in Fig. 5a $\sim 0.3/1.2=0.25$
which is a very low number in conventional optics. It is also much lower
than the SC refractive lens device reported by Cervera \textit{et al.\cite%
{FCerveraPRL02} }and the optimized homogeneous cluster of cylinders in Fig.
3c, which have a corresponding F-number close to one$.$ One explanation for
this is that the basic physics of the functionality is different leaving the
homogenization approach were the shape of the lens is the controlling
parameter, and instead using every single scatterer for bending the sound
towards the focus maximizing the amplification for the set frequency.

To see how the result holds for other frequencies, Fig. 8 shows the sound
amplification for eight different structures designed to maximize the
pressure at the same focus (located at $x=0.5$ m on the symmetry axis) for
eight different frequencies, from 300Hz to 1700Hz. There is a peak in the
frequency spectra confining the amplification for a small range of
frequencies near the optimized (indicated with a diamond symbol in the
figure). In other words, a GA lens designed to focus a given frequency is
not robust against small variations of the optimized frequency. A
substantial decrease in the amplification for lower frequency is also
observed. This result is physically understood since for larger wavelength
the cluster of cylinders appear more homogeneous to the acoustic incident
wave and the GA looses its capability of controlling and directing the
incident acoustic wave towards the focus using single scatterers.

The previous result motivated the following question: Can we create a device
that amplify the sound equally over a wide range of frequencies? This
inverse problem can be solved by modifying the fitness function in such a
way to include this condition in its definition. In what follows, we will
see how to design a device working in the range from 1000 Hz to 2000 Hz.
Thus, we have selected 6 frequencies in the range and have modify the
fitness function in the optimization process accordingly. A word of caution,
a homogenous amplification over the frequency spectra is not guaranteed if
the average pressure of the six frequencies is simply used as the fitness
for the device. To meet the homogeneous amplification criterion one has to
include the standard deviation of the six measurements. The fitness is then
calculated using the following expression

\bigskip

\begin{equation}
fitness=\overset{\_}{p}\left( 1-s^{\alpha }\right) ,
\end{equation}
where

\begin{equation}
\overline{p}=\frac{\sum_{i}|p(\upsilon _{i})|}{N},s=\sqrt{\underset{i=1}{%
\overset{N}{\sum }}\left( \left\vert p(\upsilon _{i})\right\vert -\overline{p%
}\right) ^{2},}
\end{equation}%
\noindent and $\alpha $ is a positive constant. Other alternative fitness
functions could be equally used.

By multiplying the average pressure with one minus the standard deviation we
will end up with device that will not only max out the pressure at the focus
but will have similar amplification for all six frequencies. The $\alpha -$%
constant is a factor that controls the strength of the second condition. The
time needed to calculate this design is six times greater since the SMST is
called six times \ for each fitness evaluation, one time for each frequency.
The inset of Fig. 8 shows the resulting device for $\alpha =4$. The
corresponding amplification spectrum (circles) demonstrated that the maximal
amplification is lowered from 8 dB to 4 dB, but a relative smooth
amplification is accomplished in the requested range.

\subsubsection{Nine layers lenses}

To improve the results achieved for the 5 layers flat lenses we increased
the number of cylinders in the SC structure. Figure 9 plots the
amplification for six different GA-designed 9 layers lenses with different
focal distances. In order to compare with the previous results for thinner
lenses, four extra layers were added to the left of the SC structure; in
this way the distance between the last layer and the focus is held constant.
The comparison of Figs. 7 and 9 shows that an enhancement of approximately
2-3 dB in the amplification is obtained when introducing the extra rows of
cylinders. This result gives an indication that very efficient lenses can be
designed by increasing the number of scatterers.

The multifrequency focalization device was also designed with 9 layers thick
clusters. The resulting device is shown in Fig. 10 together with its
amplification in the range 1000 Hz-2000 Hz. Notice that an enhancement in
the amplification of $\sim 1-2$ dB is obtained in comparison with the
5-layers thick device. Even if the six frequencies have a very leveled
amplification we can observe oscillations in the spectra between the
optimized frequencies used in the fitness function. These oscillations could
be eliminated by including more frequencies in the GA optimization.

Some characteristics are remarkable when the multifrequency lenses having 5
and 9 layers are compared (see Figs. 8 and 10) . First of all, the central
parts of the devices are similar. They are of triangular shape with one edge
directed at the focus and does not contain any point defects. This gives an
indication that a crystalline cluster, free from point defects, gives rise
to amplifications in a broader frequency range. And secondly, the first
layers (i.e., the ones that impinges the sound first) of the central cluster
mostly contain the cylinders having smaller radius while the triangular
cluster is mostly made of the thickest cylinders. This characteristic
results in a lower reflectance at the interface and, consequently, the
transmission across the cluster is enhanced.

\bigskip

\subsection{Robustness}

An issue of crucial interest is to know if the designed device is robust
against changes of its parameters, such as fluctuations in the radius of the
cylinders or small displacements of their position in the lattice. It has
earlier been shown \cite{ZhenYePRE03} that the focalization of sound using
homogenous 2D SC clusters depends on the displacement of the cylinders. Here
we analyzed, as a typical example, the lens shown in Fig. 5c, a flat 5 layer
lens having the focus at \textit{x}=0.5 m for an impinging wave of 1700Hz.

Table \ref{Table1} shows that the amplification decreases when the error in
the cylinders' radius increases. The amplification is calculated for five
different degrees of relative error, from 3\% to 60\%. Twenty structures was
generated and the maximal, minimal, mean and standard deviation of these is
printed out in the table. For small errors we surprisingly have a small
raise in the amplification for some structures and for a relative error of
60\% we still have solutions that conserve 100\% of the pressure in the
focus, but a mean value of 69\% for the 20 structures is calculated.

The other important issue is the sensibility of the design to changes in the
cylinders' positions. Table \ref{Table2} gives the amplification for five
different levels of random displacements in the lattice. The cylinders are
randomly moved from their original position in the lattice with the relative
error as 1\% to 15\% of the lattice parameter. The quality of the device
shows a rather high robustness since more than 90\% of the amplitude of the
pressure in the focus is conserved for a relative displacement of 12\% with
respect to the lattice parameter.

\section{Summary}

In this work we have examined how to solve the inverse problem of designing
acoustic lenses for airborne sound by using an optimization algorithm
together with a multiple scattering theory. We have introduced a symmetric
multiple scattering method to speed up the search of solutions. We have
presented different devices; e.g. thick lenses based on refractive effects
and flat lenses based on multiple scattering phenomenon. Two lenses with
flat surfaces were also designed to focus multifrequency sound waves in the
range 1 kHz to 2 kHz. The robustness were also studied against either small
variations of the cylinders' radius as well as displacements of their
positions. We hope that these predictions motivated future experimental work
that confirm the simulations. Let us stress that the procedures here
introduced can be employed to design efficient sound shields, sound filters
to a predetermined range of frequency, and many others acoustic devices for
airborne sound. Also, its application in the range of ultrasounds seems very
promising to get improved ultrasonic devices.

\newpage 
\begin{table}[tbp]
\centering%
\begin{tabular}{|c|p{2cm}|p{2cm}|p{2cm}|p{2cm}|}
\hline
$\Delta $r/r(\%) & Max & Min & Mean & S.d. \\ \hline
3 & 1.01 & 0.99 & 1.00 & $<$0.01 \\ \hline
6 & 1.02 & 0.97 & 0.99 & 0.01 \\ \hline
12 & 1.06 & 0.95 & 0.99 & 0.02 \\ \hline
20 & 1.02 & 0.88 & 0.95 & 0.04 \\ \hline
40 & 0.97 & 0.65 & 0.84 & 0.08 \\ \hline
60 & 1.00 & 0.11 & 0.69 & 0.18 \\ \hline
\end{tabular}%
\caption{Robustness of the lens in Fig. 5c against errors ($\Delta $r) of
the cylinders' radius r. Twenty random structures were calculated to do a
statistical representation. The columns are (from left to right); The
relative error with respect to the cylinders' radius; The maximum
amplification among the 20 calculated; The minimal amplification; The mean
value; The standard deviation. All expressed as parts of the amplification
for the perfect structure (8.3 dB).}
\label{Table1}
\end{table}

\bigskip

\begin{table}[tbp]
\centering%
\begin{tabular}{|c|p{2cm}|p{2cm}|p{2cm}|p{2cm}|}
\hline
$\Delta (\overrightarrow{r}-\overrightarrow{r^{\prime }})/a$ (\%) & Max & Min
& Mean & S.d. \\ \hline
1 & 1.01 & 1.00 & 1.00 & 0.00 \\ \hline
3 & 1.00 & 0.98 & 0.99 & 0.01 \\ \hline
6 & 1.00 & 0.95 & 0.98 & 0.01 \\ \hline
9 & 1.00 & 0.90 & 0.95 & 0.02 \\ \hline
12 & 0.97 & 0.85 & 0.92 & 0.03 \\ \hline
15 & 0.95 & 0.79 & 0.88 & 0.04 \\ \hline
\end{tabular}%
\caption{Robustness of the lens in Fig. 5c against errors ($\Delta (\protect%
\overrightarrow{r}-\protect\overrightarrow{r^{\prime }})$) in the
displcement of the cylinder. Twenty random structures were calculated to do
a statistical representation where each cylinder was randomly placed within
a circle, centered at the lattice point, with a radious $\Delta (\protect%
\overrightarrow{r}-\protect\overrightarrow{r^{\prime }})/a$. The columns are
(from left to right); The relative error with respect to the lattice
parameter; The maximum amplification among the 20 calculated; The minimal
amplification; The mean value; The standard deviation. All expressed as
parts of the amplification for the perfect structure (8.3 dB).}
\label{Table2}
\end{table}

\newpage

\begin{figure}
\caption{\label{fig1b}
Schematic diagram showing a
symmetric configuration of cylinders. The x-axis is the symmetry axis and
the coordinates used in the SMST are shown in the figure. The symmetric
cylinders are marked with a '+' super index.}
\end{figure}

\begin{figure}
\caption{\label{fig02}
Diagram showing that the Genetic Algorithm converges for this 100 one-bit genes problem. The curves
(corresponding scale on the left $y$-axis) correspond to the maximum (a) and
averaged (b) fitness of the population. The grey bars (corresponding scale
on the right $y$-axis) correspond to the standard deviation of the genotypes
in the population and the black bars indicate the change in the genotype
when a new maximum is found. The standard deviation is measured as the
difference in the binary digit string corresponding to a chromosome.}
\end{figure}

\begin{figure}
\caption{\label{fig3}
(color) (left panels) Four
thick SC-lenses designed for 1700 Hz and different focal distances: (a) 0.65
m, (b) 0.85 m, (c) 1.05 m, and (d) 1.25 m. (right panels) Sound
amplification in dB. The color scale used goes from blue (attenuation or dB
less than 0) to green (low aplification) to red (high amplification). The
maximum amplification in the focus is indicated in each plot.}
\end{figure}

\begin{figure}
\caption{\label{fig04}
(Left panels) Three
thick SC-lenses designed to focus the sound at $x=1.05m$ with three
different symmetry conditions (see text). The symmetry constrain is hardest
for (a) and successively lifted for (b) and (c). (right panels) Sound
amplification in dB. The color scale used goes from blue (attenuation or dB
less than 0) to green (low aplification) to red (high amplification). The
maximum amplification in the focus is indicated in each plot.}
\end{figure}

\begin{figure}
\caption{\label{fig05}
(color) (Left panels) Four flat
acoustic lenses, 5-layers-thick, designed for 1700 Hz and for four different
focal distances a) 0.3 m, b) 0.4 m, c) 0.5 m and d) 0.7 m. Three different
cylinder radius are used in the corresponding clusters: 1 cm, 1.5 cm, and 2
cm. (Right panels) Sound amplification in dB. The color scale used goes from
blue (attenuation or dB less than 0) to green (low aplification) to red
(high amplification). The maximum amplification in the focus is indicated in
each plot.}
\end{figure}

\begin{figure}
\caption{\label{fig06}
(color) (Left panels) Four flat
acoustic lenses, 5-layers-thick, designed for 1700Hz and for four different
focal distances a) 0.9 m, b) 1.1 m, c) 1.3 m and d) 2.0 m. Three different
cylinder radius are used in the cluster; 1 cm, 1.5 cm and 2 cm . (Right
panels) Sound amplification in dB. The color scale used goes from blue
(attenuation or dB less than 0) to green (low aplification) to red (high
amplification). The maximum amplification in the focus is indicated in each
plot.}
\end{figure}

\begin{figure}
\caption{\label{fig07}
Sound amplification along the
symmetry axis (\textit{y}=0) for 5-layers-thick lenses designed to focus the
sound at a given focal distance, see inset. Eight different acoustic lenses
designed for 1700 Hz are shown. Three different cylinder radiuses are used
in the corresponding lens; 1 cm, 1.5 cm, and 2 cm. The diamonds mark the 
\textit{x}-coordinate used in the fitness calculation in the GA-run.}
\end{figure}

\begin{figure}
\caption{\label{fig08}
Sound amplification (in dB) as
a function of frequency for lenses designed to focus the sound at position
x=0.5 on symmetry axis. (right panel) The continuous lines show the behavior
of eight different lenses, each one designed to focus the sound of a given
frequency, which is indicated by a diamond symbol. The circles shows the
behavior of an acoustic lens designed to focus a sound wave having any
frequency in the range from 1000 Hz to 2000 Hz. The full circles mark the
frequencies used in the fitness calculation (see Eq. (11)) (left panel) The
acoustic lens, which is made of three different sized cylinders (r=1 cm,
r=1.5 cm and r=2 cm).}
\end{figure}

\begin{figure}
\caption{\label{fig09}
Sound amplificacion along the
axis of symmetry for 9-layers-thick lenses. Eight acoustic lenses designed
for 1700Hz are shown. The symbol on each curve marks the focal distance used
in the fitness calculation in the GA and correspond to the distants in the
inset.fig9b}
\end{figure}

\begin{figure}
\caption{\label{fig10}
(right panel) Sound attenuation
as a function of frequency for an acoustic lens designed to focus the sound
at position $x=0.5m$ on the symmetry axis for any frequency in the range
from 1000 Hz to 2000 Hz. The black dots mark the frequencies $\protect\nu %
_{i}$ used in the fitness calculation (see Eq. (11)). (left panel) The
corresponding structure which is made of three different sized cylinders;
r=1 cm, r=1.5 cm and r=2 cm.}
\end{figure}


\begin{thebibliography}{widest-label}
\bibitem[*]{byline}  Electronic address: jsdehesa$@$upvnet.upv.es

\bibitem{Sigalas}  M. Sigalas and E. N. Economou, J. Sound Vib. {\bf 158},
377 (1992); for a review, see M. S. Kushawaha, Recent Res. Devel. Appl.
Phys. {\bf 2}, 743 (1999).

\bibitem{Nature}  R. Mart\'{i}nez-Sala, J. Sancho, J. V.
S\'{a}nchez-P\'{e}rez, J. Llinares and F. Meseguer, Nature (London) {\bf 378}%
, 241 (1995).

\bibitem{SanchezPerezPRL}  J. V. S\'{a}nchez-P\'{e}rez, D. Caballero, R. Mart\'{i}nez-Sala, C. Rubio, J. S\'{a}nchez-Dehesa, F. Meseguer, J. Llinares and F. G\'{a}lvez, Phys. Rev.
Lett. {\bf 80}, 5325 (1998).

\bibitem{ZhenYePRL}  Y. Y. Chen and Zhen Ye, Phys. Rev. Lett. {\bf 87},
184301 ( 2001).

\bibitem{CaballeroPRE99}  D. Caballero,  J. S\'{a}nchez-Dehesa, C. Rubio, R. Mart\'{i}nez-Sala, J. V. S\'{a}nchez-P\'{e}rez, F. Meseguer and J. Llinares, Phys. Rev. E {\bf 60}, R6316
(1999).

\bibitem{RubioJOLT}  C. Rubio, D. Caballero, J. V. S\'{a}nchez-P\'{e}rez, R. Mart\'{i}nez-Sala, J. S\'{a}nchez-Dehesa, F. Meseguer and F. Cervera, J. Lightwave Technol. {\bf 17},
2202 (1999).

\bibitem{CaballeroPRB01}  D. Caballero,  J. S\'{a}nchez-Dehesa, R. Mart\'{i}nez-Sala, C. Rubio, J. V. S\'{a}nchez-P\'{e}rez, L. Sanchis and F. Meseguer, Phys. Rev. B {\bf 64},
064303 (2001).

\bibitem{WMRobertson}  W.M. Robertson and W.F. Ruby III, J. Acoust. Soc. Am. 
{\bf 104}, 694 (1998).

\bibitem{LSanchisJASA}  L. Sanchis, F. Cervera, J. S\'{a}chez-Dehesa, J. V. S\'{a}nchez-P\'{e}rez, C. Rubio, and R. Mart\'{i}nez-Sala, J. Acoust. Soc. Am. {\bf 109}%
, 2598 (2001).

\bibitem{KushwahaAPL97}  M. S. Kushwaha, Appl. Phys. Lett. {\bf 70}, 3218
(1997)

\bibitem{GoffauxAPL03} C. Goffaux, F. Maseri, J.O. Vasseur, B. Djafari-Rouhani and Ph. Lambin, Appl. Phys. Lett. {\bf 83}, 281 (2003)

\bibitem{SanchezPerezAPL02} J. V. Sanchez-Perez, C. Rubio, R. Martinez-Sala, R. Sanchez-Grandia and V. Gomez, Appl. Phys. Lett. {\bf 81}, 5240 (2002)

\bibitem{ZhenYePRE01}  Y. Y. Chen and Zhen Ye, Phys. Rev. E {\bf 64}, 036616
(2001).

\bibitem{LSanchisPRB03}  L. Sanchis, A. H\aa kansson, F. Cervera and J. S\'{a}chez-Dehesa, Phys . Rev. B. {\bf 67},
035422 (2003).

\bibitem{FCerveraPRL02}  F. Cervera, L. Sanchis, J. V. S\'{a}nchez-P\'{e}rez, R. Mart\'{i}nez-Sala, C. Rubio, F. Meseguer, C. L\'{o}pez, D. Caballero and J. S\'{a}nchez-Dehesa, Phys. Rev. Lett. {\bf 88}
023902 (2002).

\bibitem{ZhenYePRE03}  B.C. Gupta and Z. Ye., Phys. Rev E {\bf 67}, 036603
(2003).

\bibitem{NGarciaPRE03}  N. Garcia, M. Nieto-Vesperinas, E. V. Ponizovskaya and M. Torres, Phys. Rev. E {\bf 67},
046606 (2003).

\bibitem{AAKrokhinPRL03}  A.A. Krokhin, J. Arriaga and L.N. Gumen, Phys.
Rev. Lett. {\bf 91}, 264302 (2003).

\bibitem{CHKuo04}  C. H. Kuo and Z. Ye. http://arxiv.org/abs/cond-mat/0312289 (2003).

\bibitem{AHakansson04}  A. H\aa kansson,  J. S\'{a}chez-Dehesa, F. Cervera, F. Meseguer, L. Sanchis and J. Llinares, (accepted for publication in Phys. Rev. E) http://arxiv.org/abs/cond-mat/0405104 (2003).

\bibitem{MFink00}  M. Fink, D. Cassereau, A. Derode, C. Prada, P. Roux, M. Tanter, J-L. Thomas and F. Wu, Rep. Prog. Phys. {\bf 63}, 1933
(2000).

\bibitem{MFink99}  A. Tourin, A. Derode, and M. Fink, Europhys. Lett. {\bf 47%
},175 (1999).

\bibitem{HollandBook}  J. H. Holland, Adaptation in natural and Artificial
Systems (The University of Michigan Press, Ann Arbor, 1975).

\bibitem{VTwerskyJASA51}  V. Twersky, J. Acoust. Soc. Am. {\bf 24}, 42
(1951).

\bibitem{LSanchis04}  L. Sanchis, A. H\aa kansson, D. L\'{o}pez-Zan\'{o}n, J. Bravo-Abad, and Jos\'{e} S\'{a}nchez-Dehesa, Appl. Phys. Lett. {\bf 84} 4460 (2004).

\bibitem{GoldbergBook89}  D. E. Goldberg, Genetic Algorithms in Search,
Optimization and Learning (Addison Wesley, Reading , MA, 1989).

\bibitem{GoldbergBook02}  D. E. Goldberg, The design of Innovation (Kluwer
Academic Publishers, 2002).

\bibitem{Ishimaru} Akira Ishimaru, Wave Propagation and Scattering in Random Media (Academic Press, 1978) 

\end{thebibliography}
\end{document}